\documentclass[letterpaper,12pt]{article}
\usepackage[utf8]{inputenc}
\usepackage[margin=0.75in]{geometry}
\usepackage{mathtools}
\usepackage{amsmath}
\usepackage{placeins}
\usepackage{apacite}
\usepackage{color}
\usepackage{natbib}
\usepackage{bm}

\begin{document}

\title{Statistical challenges in the analysis of sequence and structure data for the COVID-19 spike protein}

\author{Shiyu He and Samuel W.K. Wong\footnote{ Address for correspondence: Department of Statistics and Actuarial Science, University of Waterloo, Waterloo, ON, Canada.  E-mail: samuel.wong@uwaterloo.ca} \\
	Department of Statistics and Actuarial Science, University of Waterloo 
}
\date{January 30, 2021}

\maketitle

\begin{abstract}
As the major target of many vaccines and neutralizing antibodies against SARS-CoV-2, the spike (S) protein is observed to mutate over time. In this paper, we present statistical approaches to tackle some challenges associated with the analysis of S-protein data. We build a Bayesian hierarchical model to study the temporal and spatial evolution of S-protein sequences, after grouping the sequences into representative clusters. We then apply sampling methods to investigate possible changes to the S-protein's 3-D structure as a result of commonly observed mutations. While the increasing spread of D614G variants has been noted in other research, our results also show that the co-occurring mutations of D614G together with S477N or A222V may spread even more rapidly, as quantified by our model estimates.

\end{abstract}

\section{Introduction}

The severe acute respiratory syndrome coronavirus 2 (SARS-CoV-2), a strain of novel coronavirus that caused the COVID-19 outbreak in Wuhan, China in December 2019, has quickly spread across the world and has been characterized as a global pandemic \citep{ZhouP, WuF}. As of December 19, 2020, there have been over 74 million probable or confirmed cases of COVID-19, and the illness has been associated with 1.66 million deaths around the world \citep{who}.  The development of vaccines and antibody-based therapeutic agents has been initiated since the beginning of the pandemic and several have moved into phase III trials \citep{Krammer, vaccines}. Results concerning the long-term 
immunogenicity and efficacy of these vaccine candidates are a subject of continued research. Meanwhile, the virus has been found to mutate in human-to-human transmissions over time, and these changes can potentially alter the efficacy of these interventions. Therefore, it is also of vital importance to identify and study mutations with possible fitness advantages and increased infectiousness. 

SARS-CoV-2 is a single-stranded RNA virus, and RNA viruses are known to have high mutation rates and genetic diversity compared to DNA viruses \citep{Duffy, Lauring}. Their ability to evolve underlies why they can adapt to novel hosts and develop resistance to either vaccine or infection-induced immunity. Often, the most consequential mutations in terms of viral functions and resistance to neutralizing antibodies are those that alter the surface proteins of the virus. For instance, the mutation A82V in the Ebola virus glycoprotein was confirmed to have enhanced infectivity and increased the severity of the EVD epidemic \citep{Diehl}. Further, co-occurring mutations of A143V and R148K in the influenza H7N9 surface protein led to a 10-fold reduction in its sensitivity to neutralizing antibodies \citep{NingT}. As a result, mutations in the SARS-CoV-2 genome are being continuously monitored over time, and a major public repository for sequenced genomes is GISAID (\texttt{https://gisaid.org}).  In addition to collecting viral genome data, GISAID also provides tools for visualizing the spread of various mutations, organized into phylogenetic clusters (also known as \textit{clades}) over space and time.

Four structural proteins -- spike (S), envelope (E), membrane (M), and nucleocapsid (N) -- are the building blocks for the SARS-CoV-2 virus particle \citep{PhanT}. Out of these four proteins, the S-protein plays the most critical role in attachment and entry into host cells, through its binding with the human ACE2 receptor \citep{Wan}. For this reason, the S-protein is the major target of many vaccines and neutralizing antibodies against SARS-CoV-2 \citep{Amanat}. In the event of infection, these antibodies can disrupt the spike protein's ability to bind with the ACE2 receptor, thereby blocking its entry into host cells. The analyses in this paper focus on mutations in the amino acid sequence of the S-protein due to its particular importance. 

Among all currently known S-protein sequence variants resulting from mutations in the underlying genome, D614G has been the most extensively studied due to its relatively early emergence and subsequent prevalence. The notation ``D614G'' means that the amino acid D (aspartic acid) in position 614 of the original (or \textit{reference}) sequence has mutated to G (glycine), where the letters are used to denote the 20 different amino acid types. A rapid increase in D614G was observed in many regions after its initial appearance, which suggested fitness advantages and the hypothesis that variants with D614G are likely more infectious \citep{Korber}. This was later corroborated by experimental evidence that D614G, either by itself or in conjunction with other mutations, is significantly more infectious than the reference S-protein sequence \citep{LiQ}. Overall, the continued evolution of the virus has resulted in thousands of distinct S-protein sequence variants recorded in GISAID, although many of these only differ by a few mutated sequence positions. While clinical or laboratory experiments can test the infectivity of specific mutations, it is challenging to analyze large numbers of sequence variants.

Computational researchers have thus used clustering as a means to gain interpretable insight into the effect of S-protein mutations across different geographical regions.  Temporal changes in the prevalence of S-protein mutations have also been studied in related research.  For instance, \citet{chen2020mutations} clustered mutations occurring in the receptor binding domain (RBD) of the S-protein and studied binding affinity changes for each cluster. Based on common amino acid mutations, \citet{Toyoshima} classified 28 countries into three clusters and studied correlations between fatality rate and S-protein D614G variants. In addition, the hypotheses of monotonic trends for D614G and various other mutations have been tested using isotonic regression by \cite{Korber} and in the COVID-19 pipelines of the Los Alamos National Laboratory (\texttt{https://cov.lanl.gov}). However, to the best of our knowledge, few authors have built comprehensive statistical models for the evolution of S-protein mutant clusters (i.e., groups of closely related sequence variants) over space and time.  Such models can have an important practical value in providing forecasts and early warnings for countries where S-protein sequence variants with potential fitness advantages or higher infectiousness are actively being transmitted.  To that end, this paper presents one such Bayesian hierarchical model for multinomial time series that can help tackle this problem.

The 3-D structure of a protein corresponding to its amino acid sequence is a crucial part of the puzzle for understanding how the protein functions; thus, of particular interest here is the 3-D structure of the SARS-CoV-2 S-protein and its mutated variants. Often, sequence mutations associated with changes in viral infectivity can be attributed to changes in protein structure   \citep{schaefer2012predict}. The first 3-D structure of the SARS-CoV-2 S-protein was released in mid-February 2020 \citep{wrapp2020cryo}, and since then many other S-protein structures have been added to the publicly available Protein Data Bank (PDB) \citep{bernstein1977protein}. However, laboratory experiments to determine protein structure are laborious and costly, and ultimately some prove to be intractable. For this reason, the structural impacts of common S-protein mutations are not yet well-documented in the PDB, and computational methods are needed to predict their impact.  Different tools for 3-D protein structures have been used for this purpose thus far, including protein-protein binding affinity prediction \citep{chen2020mutations}, comparative modeling with known PDB structures \citep{sedova2020coronavirus3d}, and Monte Carlo sampling of protein segments \citep{wong2020assessing}. In this paper we follow the illustrative analysis in \citet{wong2020assessing}, applying similar statistical sampling approaches to assess the potential local structural changes to the S-protein for common mutations in the current mutant clusters considered. 

Overall then, our goal in this paper to illustrate statistical ideas for tackling the aforementioned challenges associated with the analysis of S-protein data, both their sequence and structure aspects. Specifically based on presently available data, we study temporal and spatial changes in the mutations of S-protein sequences and their structural impact, with the aim to better understand the ongoing evolution of the disease. Our contribution can be summarized in three parts. First, we develop a Bayesian hierarchical model to study the evolution of mutant clusters. Second, we apply sampling methods to analyze the local structural changes of the most frequently occurring protein sequence mutations in these clusters. Third, we discuss our findings and relate them to other recent work reported in the literature.

\section{Data description and exploratory analysis}

\subsection{Sequence dataset} \label{sec:sequences}

The S-protein sequence dataset for SARS-CoV-2 was obtained from GISAID on Oct 14th, 2020, with the number of sequences totaling 98,699 after incomplete sequences were removed. The full S-protein, based on the first discovered reference sequence, is 1273 amino acids long. Out of all complete sequences, 3,205 of them are unique, indicating that viral evolution has resulted in substantial genetic diversity. Our analysis of complete sequences shows that D614G is the most frequent mutation, appearing in 86.5\% of recorded sequences, followed by S477N (6.3\%), A222V (3.6\%), L18F (1.9\%), and L5F (0.99\%), R21I (0.98\%), and D936Y (0.74\%). Many of these mutations are also mentioned in the recent literature where mutation analysis was considered \citep{Korber, Chen, Hodcroft}. The sequences from GISAID are indexed by country and date of deposition, which allows us to conveniently group them for subsequent analysis.  Sequences are separately deposited by local laboratories, therefore sequence counts vary widely by country and may not be well-correlated with actual case counts. Due to this concern, we instead focus on the relative prevalence, i.e., proportions of counts, throughout this paper.

To analyze the large numbers of sequence variants, we first implemented hierarchical clustering to group the unique sequences according to their similarities. The distance matrix for hierarchical clusters was based on the number of pairwise mismatched letters and the Ward-D linkage criterion, which creates groups such that variance is minimized within clusters \citep{Ward}.  For our illustrative analysis, we chose to use a total of five clusters, which aims to achieve a balance between separability of the different clusters and interpretability of the results. Table \ref{tab:1} shows the most common mutations in each cluster and their frequencies, where it can be seen that these clusters all have identifiable patterns. For example, D614G is dominant in cluster I and present in 89\% of sequences within that cluster, while cluster II has the lowest frequency of mutations, indicating that it is composed of sequences with a high level of similarity to the original reference sequence. In clusters III, IV, and V, D614G frequently occurred together with L5F, S477N, A222V respectively, and these paired mutations are observed in almost all sequences within those clusters, evidencing a high probability for the co-occurrence of some common mutations. Evidence for these co-occurrences can also be seen in Table \ref{tab:mutpot}, which displays the top three unique sequences in each cluster (as ranked by frequency within that cluster) and their specific mutation positions.

\begin{table}[htbp]
\begin{tabular}{lllllllllll}
\hline
\text{Rank} & \multicolumn{2}{c}{\text{Cluster I}} &  \multicolumn{2}{c}{\text{Cluster II}} & \multicolumn{2}{c}{\text{Cluster III}} & \multicolumn{2}{c}{\text{Cluster IV}} &  \multicolumn{2}{c}{\text{Cluster V}} \\
& Mutation & Freq & Mutation & Freq & Mutation & Freq & Mutation & Freq & Mutation & Freq \\
\hline
1 & D614G & 89\% & P863H & 3\% & L5F & 98\% & D614G & 100\% & D614G & 99\%   \\
2 & S477N & 3\% & A262T & 2\% & D614G & 87\% & S477N & 100\% & A222V & 98\%   \\
3 & L5F & 2\% & Y453F & 2\% & H655Y & 15\% & T632N & 8\% & L18F & 54\%   \\
4 & D936Y & 2\% & T572I & 1\% & A222V & 13\% & L822F & 4\% & A262S & 13\%   \\
5 & A222V & 2\% & V615I & 1\% & V3G & 10\% & S939F & 4\% & P272L & 9\%   \\
6 & R21I & 2\% & K77M & 1\% & D574Y & 10\% & W258L & 4\% & D1163Y & 9\%   \\
7 & L54F & 1\% & A845S & 1\% & S459Y & 6\% & E1144Q & 4\% & L5F & 7\%   \\
8 & P1263L & 1\% & L8V & 1\% & M1229I & 4\% & G566S & 4\% & G1167V & 6\%   \\
9 & Q677H & 1\% & H655Y & 1\% & T859I & 4\% & P330A & 2\% & L176F & 4\%  \\
\hline
\end{tabular}
\caption{Relative frequencies of the most common mutations present in each cluster. The top 9 mutations in descending order and their corresponding relative frequencies are shown in the columns for each of clusters I--V.  For example, the D614G mutation is present in 89\% of the sequences belonging to cluster I.} 
    \label{tab:1}
\end{table}

\begin{table}[ht]
\begin{center}
\begin{tabular}{crl}
\hline
Cluster & Frequency & Mutation Positions   \\
\hline
I & 58,271 & D614G   \\
  & 777 & R21I, D614G   \\
  & 602 & D614G, D936Y   \\\hline
II & 11,617 & Reference Sequence   \\
  & 117 & A829T   \\
  & 43 & L8V   \\\hline
III & 486 & L5F, D614G   \\
  & 121 & L5F   \\
  & 43 & L5F, A222V, D574Y, D614G, H655Y   \\\hline
IV & 5,472 & S477N, D614G   \\
  & 85 & S477N, D614G, T632N   \\
  & 51 & S477N, D614G, A930V   \\\hline
V & 1,388 & A222V, D614G   \\
  & 1,369 & L18F, A222V, D614G   \\
  & 150 & A222V, A262S, P272L, D614G \\
\hline
\end{tabular}
\caption{Top three unique sequences in each cluster, ranked by frequency.
For each unique sequence, its specific mutation positions are shown in the right column.
For example, 58,271 sequences belonging to cluster I had exactly the one mutation D614G, while 777 sequences in cluster I had exactly the two mutations R21I and D614G.} \label{tab:mutpot}
\end{center}
\end{table}

\begin{figure}[htbp]
\begin{center}
\includegraphics[width = \linewidth]{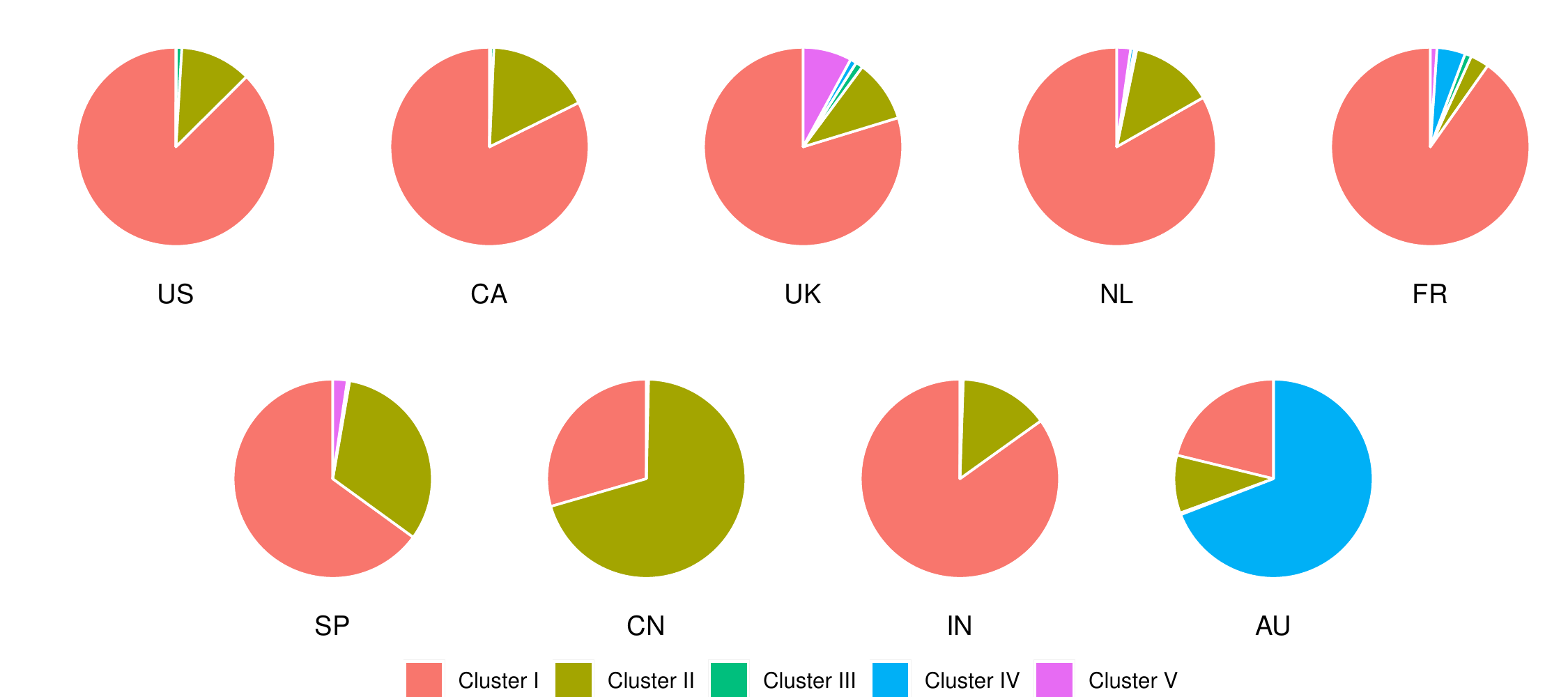}
\caption{Pie charts of cluster proportions for 9 countries, based on all complete sequences accumulated in GISAID. For each country, the proportions of cluster I to cluster V are represented by the 5 different colors as indicated.}
\label{fig:piechart}
\end{center}
\end{figure}

We selected 9 countries from the GISAID database to study based on the larger numbers of sequences deposited, which are United States (US), Canada (CA), United Kingdom (UK), Netherlands (NL), France (FR), Spain (SP), China (CN), India (IN), and Australia (AU). The pie charts in Figure \ref{fig:piechart} show the composition of clusters for each country for sequences accumulated since the outbreak. Countries with distinctly different compositions are China and Australia, where China has the majority of its sequences from cluster II, and Australia has cluster IV as its major cluster. Cluster I is the largest cluster for the rest of the countries, followed by cluster II, while cluster V has a noticeable presence in Europe, especially the UK.

\begin{figure}[htbp]
\begin{center}
\includegraphics[width = 13cm]{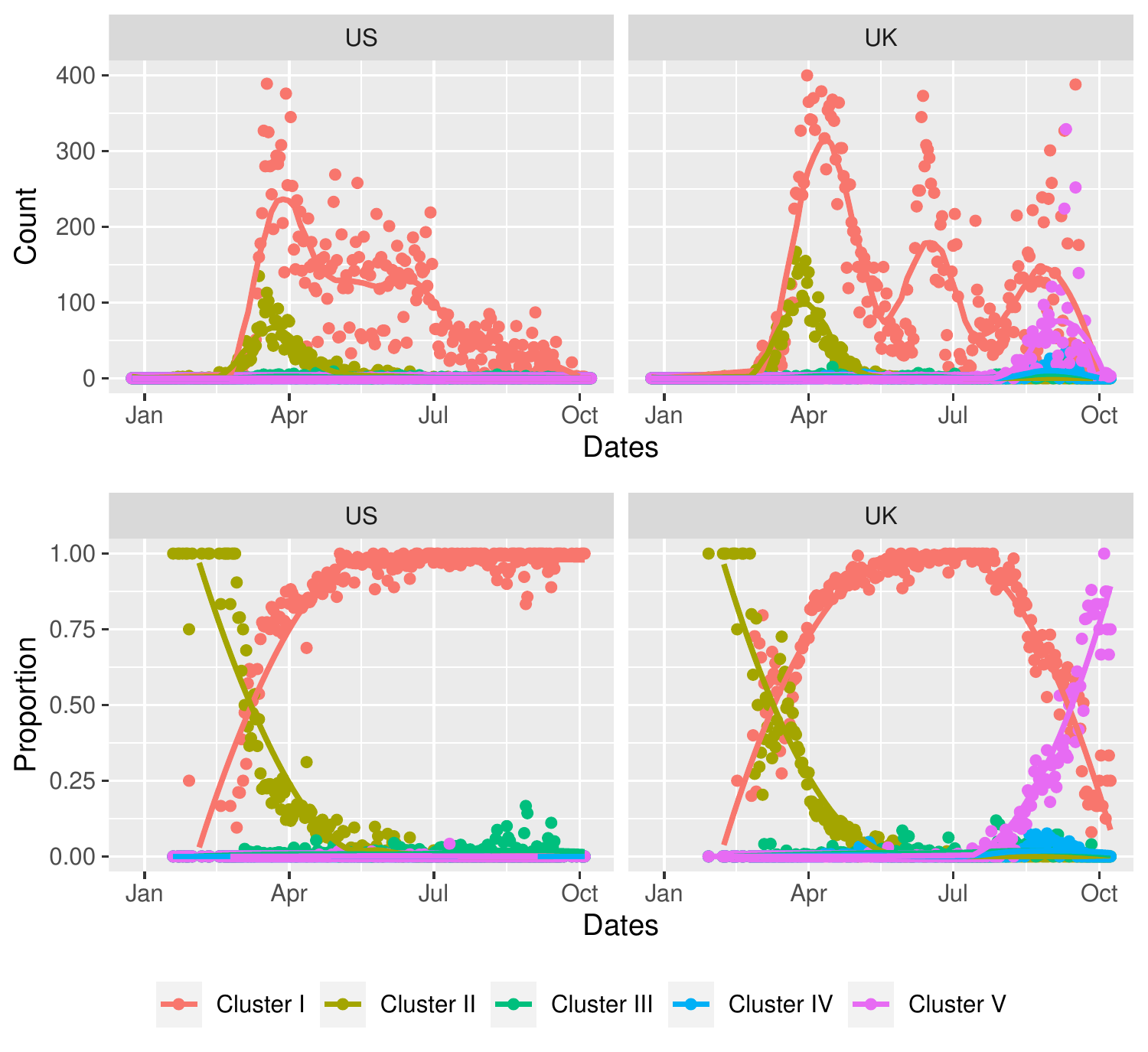}
\caption{Temporal trend of cluster counts (upper panels) and cluster proportions (lower panels) in the US and UK.  The points indicate the observed proportions and counts on each day, while the solid lines show the corresponding smoothed LOESS curves.}
\label{fig:clusterUSUK}
\end{center}
\end{figure}

The composition of clusters also changes over time, and as examples we show the temporal trend of the daily cluster counts (upper panels) and proportions (lower panels) for the United States (US) and the United Kingdom (UK) in Figure \ref{fig:clusterUSUK}. The graphs show a major difference between the US and UK composition trends over time:  cluster I in the UK peaked around May to July and cluster V saw a surge since late August or early September, while cluster I remains dominant in the US.  Similarly, most European countries have cluster V surging during this period, which suggests the most common mutations in cluster V, D614G and A222V, might be related to the rise of infections across Europe during the late summer and early autumn of 2020 \citep{DongE}. In addition to this, the majority of countries have cluster II proportions decreasing over time in direct correspondence to the emergence of cluster I infections over time.  It can be seen that the GISAID sequence counts in the upper panels of Figure \ref{fig:clusterUSUK} vary widely over time and do not necessarily correspond well with the actual case counts in the US and UK during this period.

\subsection{Structure dataset}\label{sec:structuredata}

The 3-D structure data for the S-protein were obtained from the Protein Data Bank (PDB) \citep{bernstein1977protein}.  The first laboratory-determined of a standalone 3-D structure of the SARS-CoV-2 S-protein was contributed in mid-February 2020 by a team of scientists at UT Austin using cyro-EM techniques \citep{wrapp2020cryo}, with PDB accession code 6VSB.  Since then, many other groups around the world have contributed to the effort of studying different aspects of the S-protein using laboratory techniques.  As of Oct 14th, 2020, there were 108 3-D structures publicly available in the PDB associated with the SARS-CoV-2 S-protein:  40 of these considered the S-protein in isolation, under different conformational states and sequence variants; 11 of these studied the structure of S-protein when bound together with ACE2; the remaining 57 studied the structure of S-protein when interacting with different potential antibodies.  Of the 68 structures containing the S-protein bound together with ACE2 or an antibody, 32 of these focused on a specific region of the S-protein, known as the receptor binding domain (RBD), primarily to decipher the binding behaviour of the S-protein to these molecules.  Otherwise, for the majority of the structures (76 out of 108), experimenters attempted to determine the structure for the full S-protein.

\begin{figure}[h]
\includegraphics[width = \textwidth]{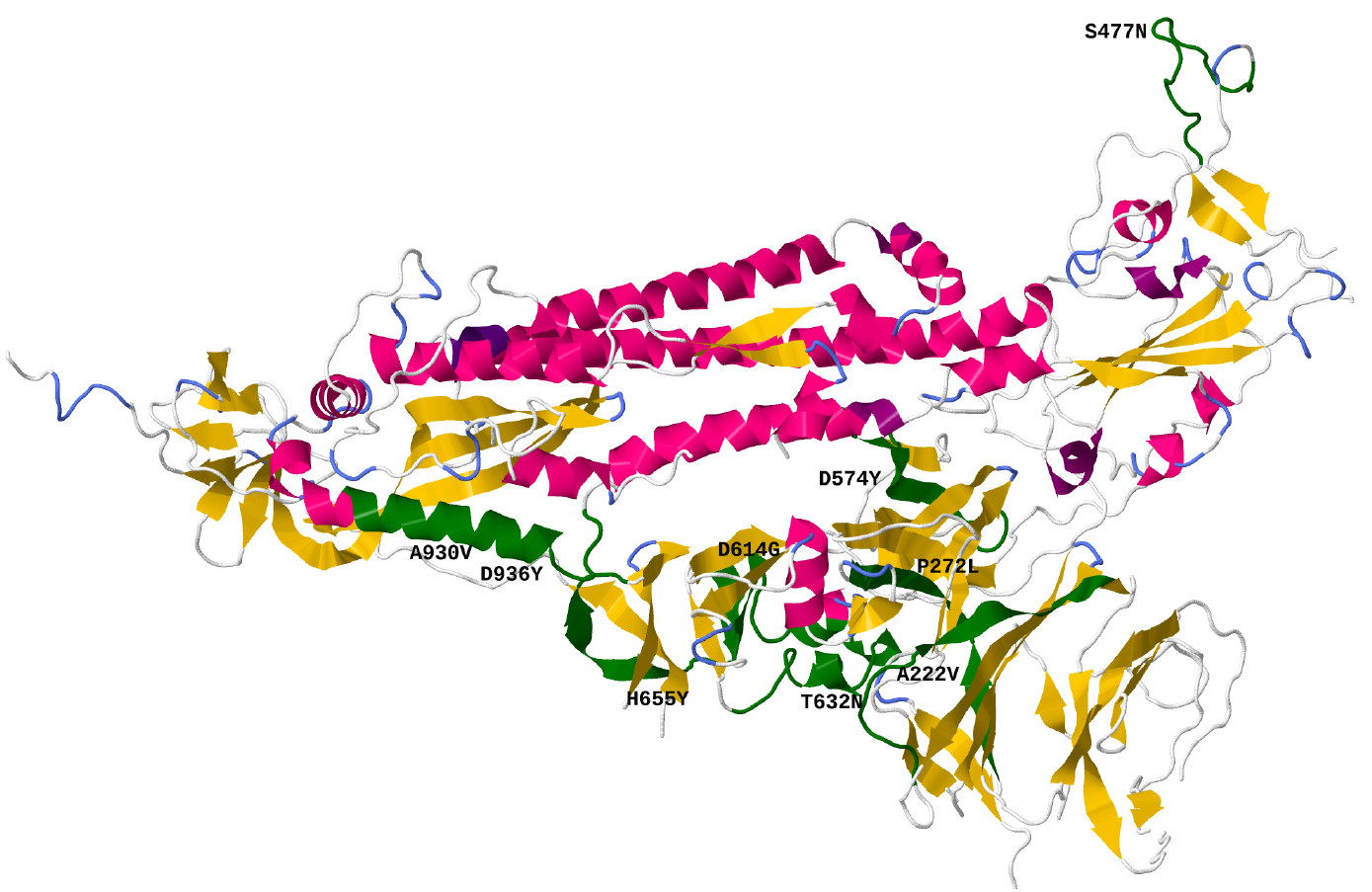}
\caption{3-D structure of the reference sequence S-protein (PDB accession code 6XM0). Protein segments containing common mutations are highlighted in green and labeled with the corresponding mutation in Table \ref{tab:mutpot};  segments with mutations that have incomplete 3-D structure data are not shown.}
\label{fig:ref3D}
\end{figure}

Together, the PDB reflects the current state of knowledge for the S-protein structure, including the attempts made to assess possible neutralizing antibodies in the development of therapeutic interventions.  Overall, 3-D structure determination has not kept pace with genome sequencing: among all the amino acid mutations listed in Table \ref{tab:1}, only D614G has been studied in the laboratory.  Thus, we cannot leverage the PDB to ascertain the potential changes in the 3-D structure of the S-protein as a result of those mutations. Even in the best consensus 3-D structure for the reference sequence to date \citep{zhou2020cryo}, parts of the S-protein have not been successfully determined by the laboratory experiments, resulting in missing data.  This consensus structure (PDB accession code 6XM0) is visualized in Figure \ref{fig:ref3D}, where the  mutations identified in Table \ref{tab:mutpot} are labelled; mutation locations where there is no 3-D structural information available are omitted from the figure.

\section{Methods and models}

\subsection{Bayesian hierarchical model for mutant clusters of sequences}\label{sec:methodseq}

In this section we present a Bayesian hierarchical model for the observed sequence counts, by day and country, in each of the five clusters identified via our exploratory analysis.  In motivating the model, we recall that the number of deposited GISAID sequences varies widely over time and by country.  Thus, it is sensible to focus inference on the fraction of sequences belonging to each cluster (i.e., cluster proportions) in each country, as in compositional data analysis \citep{aitchison1982statistical}. Also, while LOESS identified some temporal trends for these cluster proportions in each country, it cannot be used for prediction or for comparing the overall growth rates of different clusters. Thus, our model contains the following key features.  First, it provides estimates of the cluster proportions for each country on any given day.
Second, it assumes a growth rate parameter for each cluster that is common across all countries, through which differences in prevalence among the S-protein mutant clusters can be quantified. Third, the temporal evolution of the cluster proportions  is allowed to be dependent across countries via correlated errors.

Denote the observed sequence counts by the vectors $\bm{y}_{it} = (y_{it1}, \ldots, y_{it5} )$, where $y_{itc}$ is the number of sequences observed in GISAID for country $i = 1, \ldots, K$ on day $t \ge 1$ belonging to  cluster $c = 1, \ldots 5$ (corresponding to clusters I--V in section \ref{sec:sequences}). We then let
\begin{eqnarray}
\bm{y}_{it} \sim 
Multinomial \left( n_{it}, \frac{1}{14} \sum_{j=t-13}^t \bm{p}_{ij} \right) \mbox{ for } t \ge 14 \nonumber
\end{eqnarray}
where $n_{it} = \sum_{c=1}^5 y_{itc}$ is the total number of GISAID sequences recorded for country $i$ on day $t$, and $\bm{p}_{it} = (p_{it1}, \ldots, p_{it5} )$ is the vector of probabilities representing the true underlying cluster proportions for country $i$ on day $t$ such that $ \sum_{c=1}^5 p_{itc} = 1$. Thus the model assumes that the observed sequences represent a random sample from the population of infected individuals in a country, with a reporting delay uniformly at random over the commonly assumed 14 day incubation period for the virus \citep{lauer2020incubation}. The sampling fraction (i.e.,  number of reported GISAID sequences out of the number of infected individuals in the country) may  vary over time, and this setup permits inference on   $\bm{p}_{it}$ in a manner that accommodates that sampling variability.  Prior to Jan 20th, 2020 (i.e., when $t < 14$ in the model), we simply take the multinomial probability to be the average of the days modeled thus far, namely  $\frac{1}{t} \sum_{j=1}^t \bm{p}_{ij}$.

We apply a log-ratio transformation on the cluster proportions, as commonly used in compositional data analysis \citep{aitchison1999logratios}.  Treating cluster I as the baseline, we define
\begin{eqnarray}
\tilde{\bm{p}}_{it} = \left[\log  \left( \frac{p_{it2}}{p_{it1}} \right), \log  \left( \frac{p_{it3}}{p_{it1}}\right), \log   \left(\frac{p_{it4}}{p_{it1}}\right), \log  \left(\frac{p_{it5}}{p_{it1}}\right)\right] \nonumber
\end{eqnarray}
and model these transformed $\tilde{\bm{p}}_{it}$ values according to
\begin{eqnarray}
\tilde{\bm{p}}_{it} = \tilde{\bm{p}}_{i,t-1} + \bm{\alpha} + \bm{\epsilon}_{it}  \mbox{ for } t > 1  \label{eq:dynmodel}
\end{eqnarray}
where $\bm{\alpha} = (\alpha_2, \alpha_3, \alpha_4, \alpha_5)$ is a vector of growth rate parameters for clusters II to V, and $\bm{\epsilon}_{it} = (\epsilon_{it2}, \epsilon_{it3}, \epsilon_{it4}, \epsilon_{it5} )$ represents a random noise vector with mean zero that governs daily fluctuations.  Note that when   $\bm{\epsilon}_{it} = \bm{0}$, equation (\ref{eq:dynmodel}) implies
\begin{eqnarray}
p_{itc} = \frac{p_{i,t-1,c} \exp(\alpha_c)}{ \sum_{c=1}^5 {p_{i,t-1,c} \exp(\alpha_c)}} \propto p_{i,t-1,c} \exp(\alpha_c) \nonumber
\end{eqnarray}
for $c=1, \ldots, 5$ with $\alpha_1$ defined to be 0, so that $\exp (\bm{\alpha})$ can be interpreted as the multiplicative daily growth rates of clusters II to V relative to cluster I, with the proportions normalized to sum to 1, similar to a model recently used to study the prevalence of different flu strains \citep{huddleston2020integrating}. The $\bm{\alpha}$ are assumed to be the same for all countries, to reflect the intrinsic fitness of each mutant cluster.  Define $\bm{\epsilon}_{1:K, t, c} = ( \epsilon_{1tc}, \ldots, \epsilon_{Ktc})$, namely the random noise vector on day $t$ across all $K$ countries for cluster $c$, where $c \in \{2,3,4,5\}$.  Then we let $\bm{\epsilon}_{1:K, t, c} \sim N_K (\bm{0}, \Sigma)$ independently for each day and cluster, where $N_K$ denotes a $K$-variate Normal distribution and $\Sigma$ is a covariance matrix.  This formulation allows for spatial dependence in the sense that the noise term can be correlated among countries; intuitively, if a certain cluster experiences faster than expected growth in one country for a period of time, nearby or geographically linked countries may experience similar changes.  In practice, we could set $\Sigma$ to be of block diagonal form, for example, with each block as countries within the same continent, assuming correlations between different continents are likely negligible.  Overall, our model setup follows the general structure of Gaussian dynamic models for multinomial time series described in \citet{cargnoni1997bayesian}.

We complete the model specification with the choice of priors.
During the early outbreak period from Dec 24th, 2019 to Jan 6th, 2020, a total of 40 sequences worldwide were deposited in GISAID, with two that we would now classify to be in cluster I and 38 in cluster II.  Thus we set a Dirichlet prior for the cluster proportions in the model for the starting date of the model (Jan 7th, 2020),  $\bm{p}_{i1} \sim Dirichlet(3, 39, 1, 1, 1)$ independently for all $i=1,\ldots K$ countries, obtained by adding the cluster I and II observations as pseudocounts to a uniform Dirichlet distribution.
Weakly informative Cauchy priors with scale 0.5 are assigned to the each unique diagonal (variance) element of $\Sigma$. A weakly informative LKJ correlation distribution with shape parameter 2 is assigned as the prior for the correlation matrices corresponding to the blocks of $\Sigma$. Finally, uniform priors are assigned for $\bm{\alpha}$.

To obtain the samples for the posterior distribution of the parameters, Markov chain Monte Carlo sampling for the model was carried out via Stan \citep{carpenter2017stan}.  Four parallel chains were run, with 5000 iterations each and the first half discarded as burn-in.

\subsection{Sampling methods for local protein structure analysis}

Certain mutant clusters may have a higher prevalence than others, as identified via the estimates for $\bm{\alpha}$.  A natural follow-up question is to ask whether these differences might be related to changes in the 3-D structure of the S-protein as a result of the common mutations shown in Table \ref{tab:mutpot}. To compare two 3-D protein structures, it is standard practice to compute the root-mean-square deviation (RMSD) between its corresponding backbone atoms; the four backbone atoms (N, C$_\alpha$, C, O) are common to all amino acids, so this RMSD calculation can be applied even when amino acid mutations are present and provides a simple metric for assessing structural changes. However, as described in section \ref{sec:structuredata}, currently the PDB lacks laboratory-determined structures for all but the D614G mutation, and thus modeling approaches are needed.

A protein structure is represented by the arrangement of its atoms in 3-D space, which is known as a \textit{conformation}.  Letting $x$ denote a conformation and $H$ a given scalar energy (or potential) function, a statistics-based approach to the problem is to draw samples from the Boltzmann distribution
\begin{eqnarray}
\pi(x) \propto \exp \left\{-H(x)/T\right\},  \label{pix}
\end{eqnarray}
where $T$ is the effective temperature. According to the energy landscape theory \citep{onuchic1997theory}, a protein structure tends to be most stable around the lowest energy conformation.  While nature's `true' energy function is not known, various energy approximations $H$ have been developed to mimic this property for use in computational protein structure prediction \citep{zhang2007monte}.  Thus in this context, the goal is to draw samples from equation (\ref{pix}) corresponding to the amino acid sequence before and after mutation, to assess possible structure differences among the low-energy conformations sampled.  

We focus here on local structural impacts, that is, possible changes to the protein structure in the segment of amino acids near the mutation position.  To do so, we treat the PDB structure in \citet{zhou2020cryo} (with accession code 6XM0, and visualized in Figure \ref{fig:ref3D}) as the reference consensus 3-D structure for the S-protein corresponding to the reference sequence.  Then we may sample conformations for specific segments of amino acids while holding the rest of the structure fixed at the coordinates in this reference 3-D structure. Segment lengths of up to approximately 15 amino acids has been recognized to be a rough upper bound where current sampling methods can perform adequately \citep{webb2017protein}.  For an individual mutation occurring at position $j$, we thus sample conformations for the length 15 segment of amino acids from positions $j-7$ to $j+7$.   These length 15 segments for each individual mutation listed in Table \ref{tab:mutpot} are highlighted in green in Figure \ref{fig:ref3D}. 

Since proteins are composed of a linear sequence of amino acids, a sequential sampling approach can effectively exploit that property, by 
incrementally adding one amino acid at a time to construct approximate samples from equation (\ref{pix}). The idea of devising sequential sampling algorithms as a way to stochastically search for realistic low-energy conformations was originally proposed in \citet{zhang2007biopolymer} and tested on lattice representations of proteins.  Subsequently, extensions of the method applicable to real protein structures have been developed, including distance-guided chain growth \citep[DisGro,][]{tang2014fast} and sequential Monte Carlo \citep[SMC,][]{wong2018exploring}.  The implementations of these two algorithms also use slightly different approximations for the energy function $H$, and together can provide a more complete picture of the energy landscape. Thus we apply these algorithms to sample conformations for the segments shown in Figure \ref{fig:ref3D}, on both the reference sequence and the mutated sequence. Then following \citet{wong2020assessing}, we may compute the probability distribution of RMSDs between pairs of sampled conformations, as a way to assess potential differences in the low-energy conformational space as a result of the mutation. Specifically for the D614G mutation, we can use its known structure in the PDB to validate the results of these sampling methods.

\section{Results}

We present our results from fitting the proposed Bayesian hierarchical model in section \ref{sec:resultseq} and the results of sampling 3-D protein conformations in section \ref{sec:resultstruct}. 

\subsection{Estimates of growth for the mutant clusters of sequences}\label{sec:resultseq}

\begin{figure}
\begin{center}
\includegraphics[width = 0.9\linewidth]{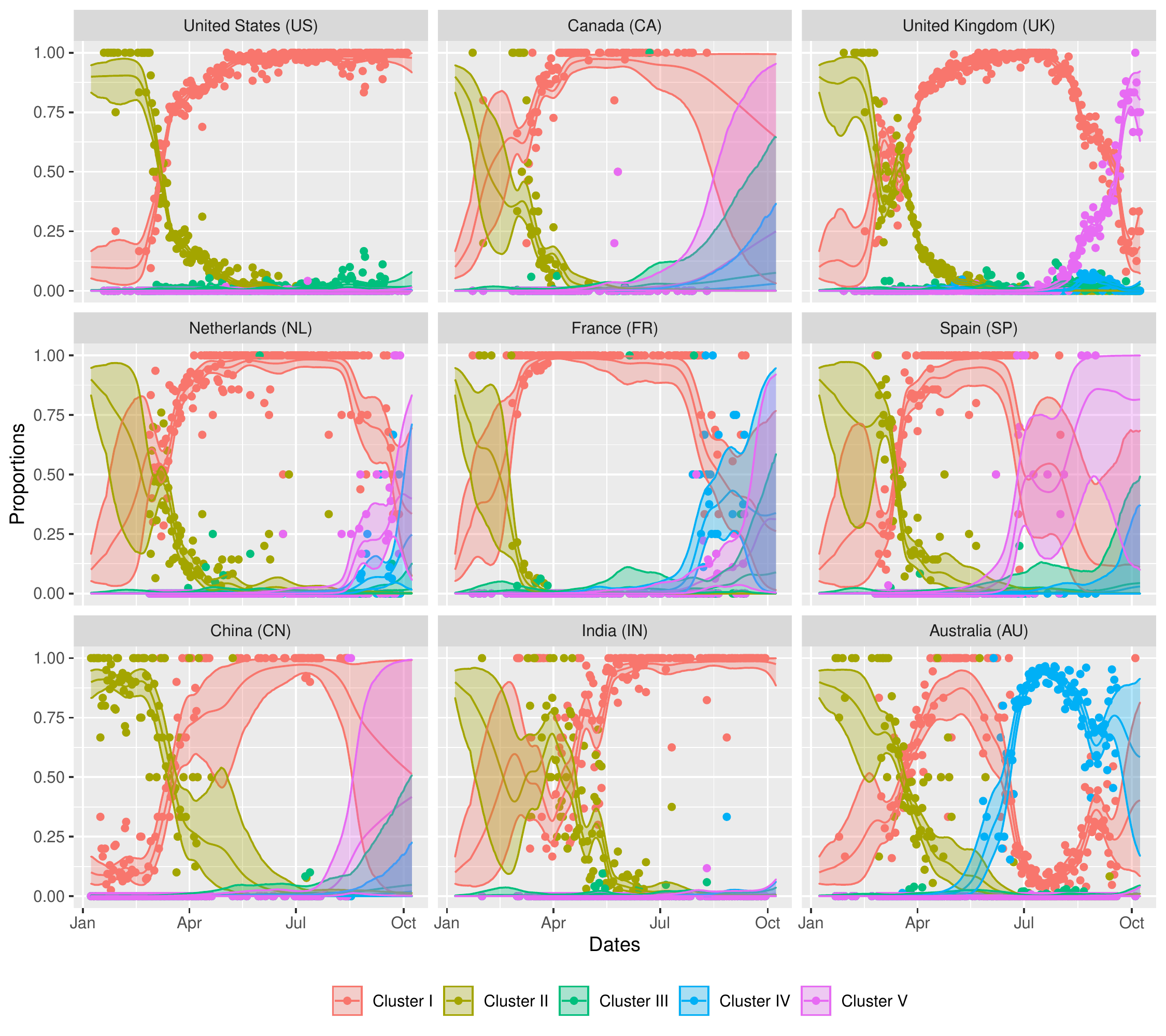}
\caption{Posterior means and 95\% credible intervals of inferred cluster proportions from the Bayesian hierarchical model for 9 countries. For each day, the points show the observed proportions, the middle solid lines indicate the posterior means of cluster proportions, and the bands indicate the 95\% credible interval of cluster proportions.} \label{fig:postclust}
\end{center}
\end{figure}

The posterior distribution of the parameters in the Bayesian hierarchical model (Table \ref{tab:3}) show that the daily growth rate parameters of clusters II to V relative to cluster I, estimated via their posterior means, are -0.05 (-0.07, -0.03), 0.00 (-0.01, 0.02), 0.02 (-0.01, 0.04), 0.03 (0.00, 0.05), with 95\% credible intervals in brackets. These estimates show that during the study period, the sequences from clusters IV and V tend to have higher growth relative to clusters I, II, and III, of which mutant cluster II clearly has the weakest growth, as seen via its posterior interval that does not overlap the others. Referring to the mutation positions in Table \ref{tab:mutpot}, this indicates that sequences with amino acid D in position 614 (as in the reference sequence) will tend to decrease in prevalence over time in the presence of the other mutant clusters.  In addition, cluster I, which is mostly characterized by the lone D614G mutation, may have a growth disadvantage if clusters IV or V are also spreading in the country. The clusters with the strongest growth (IV and V) are primarily composed of variants with the co-occurrence of D614G with S477N or A222V. 

\begin{table}[htbp]
\centering
\begin{tabular}{rrrrr}
  \hline
 Parameter & Mean & 2.5\% & 97.5\% \\ 
  \hline
  $\alpha_2$  & -0.0495 & -0.0658 & -0.0344 \\ 
  $\alpha_3$ & 0.0022 & -0.0142 & 0.0178 \\ 
  $\alpha_4$ & 0.0172 & -0.0073 & 0.0398 \\ 
  $\alpha_5$ & 0.0267 & 0.0047 & 0.0478 \\ 
  $\sigma_{\text{NA}}$ & 0.2432 & 0.1664 & 0.3820 \\ 
  $\sigma_{\text{EU}}$ & 0.3802 & 0.2915 & 0.4771 \\ 
  $\sigma_{\text{AS}}$ & 0.3578 & 0.1885 & 0.6661 \\ 
  $\sigma_{\text{AU}}$ & 0.2768 & 0.1989 & 0.3926 \\ 
  $p_{\cdot,1,1}$ & 0.0999 & 0.0524 & 0.1660 \\ 
  $p_{\cdot,1,2}$ & 0.8993 & 0.8327 & 0.9473 \\ 
  $p_{\cdot,1,3}$ & 0.0008 & 0.0001 & 0.0030 \\ 
  $p_{\cdot,1,4}$ & 0.0000 & 0.0000 & 0.0001 \\ 
  $p_{\cdot,1,5}$ & 0.0000 & 0.0000 & 0.0001 \\ 
  $\Sigma_{\text{NA} 1,1}$ & 0.0618 & 0.0277 & 0.1459 \\ 
  $\Sigma_{\text{NA} 1,2}$ & -0.0059 & -0.0475 & 0.0344 \\ 
  $\Sigma_{\text{NA} 2,1}$ & -0.0059 & -0.0475 & 0.0344 \\ 
  $\Sigma_{\text{NA} 2,2}$ & 0.0618 & 0.0277 & 0.1459 \\ 
  $\Sigma_{\text{EU} 1,1}$ & 0.1468 & 0.0850 & 0.2276 \\
  $\Sigma_{\text{EU} 1,2}$ & 0.0486 & -0.0353 & 0.1429 \\ 
   \hline
   \end{tabular}
\begin{tabular}{rrrrr}
  \hline
  Parameter & Mean & 2.5\% & 97.5\% \\ 
  \hline
  $\Sigma_{\text{EU} 1,3}$ & 0.0431 & -0.0309 & 0.1313 \\ 
  $\Sigma_{\text{EU} 1,4}$ & -0.0133 & -0.1055 & 0.0841 \\ 
  $\Sigma_{\text{EU} 2,1}$ & 0.0486 & -0.0353 & 0.1429 \\ 
  $\Sigma_{\text{EU} 2,2}$ & 0.1468 & 0.0850 & 0.2276 \\ 
  $\Sigma_{\text{EU} 2,3}$ & 0.0178 & -0.0610 & 0.1048 \\ 
  $\Sigma_{\text{EU} 2,4}$ & 0.0320 & -0.0926 & 0.1294 \\ 
  $\Sigma_{\text{EU} 3,1}$ & 0.0431 & -0.0309 & 0.1313 \\ 
  $\Sigma_{\text{EU} 3,2}$ & 0.0178 & -0.0610 & 0.1048 \\ 
  $\Sigma_{\text{EU} 3,3}$ & 0.1468 & 0.0850 & 0.2276 \\ 
  $\Sigma_{\text{EU} 3,4}$ & 0.0153 & -0.0802 & 0.1114 \\ 
  $\Sigma_{\text{EU} 4,1}$ & -0.0133 & -0.1055 & 0.0841 \\ 
  $\Sigma_{\text{EU} 4,2}$ & 0.0320 & -0.0926 & 0.1294 \\ 
  $\Sigma_{\text{EU} 4,3}$ & 0.0153 & -0.0802 & 0.1114 \\ 
  $\Sigma_{\text{EU} 4,4}$ & 0.1468 & 0.0850 & 0.2276 \\ 
  $\Sigma_{\text{AS} 1,1}$ & 0.1422 & 0.0355 & 0.4436 \\ 
  $\Sigma_{\text{AS} 1,2}$ & 0.0051 & -0.1227 & 0.1803 \\ 
  $\Sigma_{\text{AS} 2,1}$ & 0.0051 & -0.1227 & 0.1803 \\ 
  $\Sigma_{\text{AS} 2,2}$ & 0.1422 & 0.0355 & 0.4436 \\ 
  & & &\\
   \hline
   \end{tabular}   
\caption{Posterior means and 95\% credible intervals (represented by the 2.5\% and 97.5\% percentiles) for the parameters in the Bayesian hierarchical model. $\bm{\alpha}$ represents the daily growth rate parameters for clusters II to V relative to cluster I. $\sigma_C$ is the random noise standard deviation for countries within continent $C$, and $\bm{p}_{\cdot,1}$ represents the initial proportions on Jan 7th, 2020 for all countries. $\Sigma_C$ for each continent together forms the diagonal block matrix $\Sigma$ that represents the covariance matrix for the random noise vector.}
\label{tab:3}
\end{table}

The covariance matrix $\Sigma$ for the random noise vector is set up in block diagonal form:  each block represents countries within the same continent and parameterized as $\Sigma_{C} = diag(\sigma_C) \times \Omega_C \times diag(\sigma_C)$, where $\sigma_C$ is the standard deviation of the daily noise term for countries within continent $C$, and $\Omega_C$ is the corresponding correlation matrix. Specifically, we defined four continents:  North America (NA) as (US, CA); Europe (EU) as (UK, NL, FR, SP); Asia (AS) as (CN, IN); and Australia (AU) as its own continent. The posterior means for $\sigma_{\text{EU}}$ and $\sigma_{\text{AS}}$ are relatively larger than $\sigma_{\text{NA}}$ and $\sigma_{\text{AU}}$, indicating that overall cluster growth trends are somewhat more predictable in North America and Australia than Europe or Asia. On the other hand, spatial dependence in the noise terms in general is low for countries studied, with no estimated correlations exceeding 0.4. That said, based on the posterior means, European countries still have relatively larger spatial correlations in their daily fluctuations, e.g., UK and Netherlands have a correlation of 0.33, UK and France have a correlation of 0.28, and Netherlands and Spain have a correlation of 0.22. The estimates show these European countries may experience more similar day-to-day changes, while countries in North America and Asia do not, as the correlation appears to be negligible between US and Canada (-0.087) and between China and India (-0.045), both of which are close to 0. The relatively low correlations between countries suggest that zero noise correlation between continents (i.e., block diagonal $\Sigma$) is a reasonable simplifying assumption. 

Figure \ref{fig:postclust} shows the posterior means and 95\% credible intervals for the inferred cluster proportions in the different countries on each day $t$ from Jan 7th, 2020 to Oct 14th, 2020. The posterior means of $\bm{p}_{\cdot,1}$ indicate that on Jan 7th, 2020, cluster I accounts for around 10\% and cluster II accounts for around 90\% of sequences.  Nonetheless these initial proportions are quite uncertain  due to the limited number of early cases, as seen in the wide credible intervals on the plot. The credible intervals narrow as we reach periods where a larger number of sequences are deposited.   Overall, we see growth for cluster I accelerates during January to June but appears to rapidly fall off in July for many countries, while maintaining a substantive presence in the US, India and Australia. Cluster II clearly diminishes over time worldwide, and cluster III is fairly small but stable for all countries.  Cluster IV estimates show small fluctuations for most countries except Australia, where it expands to over 90\% from July to August and may remain as the dominant cluster. Cluster V is estimated to first appear in August and thereafter shows a rapid growth in all European countries.

Our Bayesian hierarchical model also allows prediction of changes in cluster proportions for countries with missing or very sparse GISAID sequence data.  Canada, Spain and China only have deposited sequences up to August, while France only has sequence data deposited until mid-September; our model is nonetheless able to provide the point and interval estimates for their cluster proportions over the entire period.  As suggested in Figure \ref{fig:postclust}, both Canada and China are projected to have cluster I gradually decrease together with a possible rise in cluster V;  the expanding credible intervals reflect the increasing uncertainty associated with the increasing number of days with missing data. Meanwhile, the main clusters present in France by mid-October might be IV or V (or their combination), while Spain is projected to be dominated by cluster V much like the rest of Europe.

To ensure that our results are robust, we performed a sensitivity analysis on the posterior parameters given different sets of priors, with a special focus on the sensitivity of $\bm{\alpha}$ and $\bm{p}_{i1}$. We created the following three scenarios to compare with our base scenario prior choices in Section \ref{sec:methodseq}, and the results are shown in Table \ref{tab:sensitivity}. In scenario 1, we set the cluster proportions on the starting date as $\bm{p}_{i1} \sim Dirichlet(2.5, 38.5, 0.5, 0.5, 0.5)$, obtained by adding the pseudocounts to Dirichlet parameters corresponding to the Jeffreys prior, that is, $Dirichlet(0.5,0.5,0.5,0.5,0.5)$.  In scenario 2, we set the priors for each growth parameter $\alpha_c \sim N(0,1)$ independently for $c=2,\dots, 5$, instead of the uniform priors in our base scenario. In scenario 3, we combined both changes to the priors made in scenario 1 and 2. Compared with the main results in Table \ref{tab:3}, all three scenarios in Table \ref{tab:sensitivity} show comparable posterior means and 95\% credible intervals with the base scenario, which indicates our posterior parameters are fairly stable and robust to the different choices of priors. 

\begin{table}[htbp]
\centering
\begin{tabular}{rrrrrrrrrr}
  \hline
  & \multicolumn{3}{c}{Scenario 1} & \multicolumn{3}{c}{Scenario 2} & \multicolumn{3}{c}{Scenario 3}\\
  Parameter & Mean & 2.5\% & 97.5\% & Mean & 2.5\% & 97.5\%  & Mean & 2.5\% & 97.5\%\\ 
  \hline
  $\alpha_2$ & -0.0492 & -0.0655 & -0.0351 & -0.0494 & -0.0656 & -0.0345  & -0.0503 & -0.0664 & -0.0352 \\ 
  $\alpha_3$ & 0.0040 & -0.0110 & 0.0188 & 0.0020 & -0.0137 & 0.0173 & 0.0040 & -0.0119 & 0.0195 \\ 
  $\alpha_4$ & 0.0240 & 0.0007 & 0.0467 & 0.0177 & -0.0069 & 0.0407  & 0.0242 & -0.0000 & 0.0468\\ 
  $\alpha_5$ & 0.0341 & 0.0127 & 0.0552 & 0.0266 & 0.0050 & 0.0470 & 0.0338 & 0.0116 & 0.0548\\ 
  $p_{\cdot,1,1}$ & 0.0839 & 0.0423 & 0.1418 & 0.1006 & 0.0528 & 0.1668 & 0.0850 & 0.0435 & 0.1447\\ 
  $p_{\cdot,1,2}$ & 0.9157 & 0.8573 & 0.9575 & 0.8986 & 0.8319 & 0.9465 & 0.9145 & 0.8546 & 0.9563\\ 
  $p_{\cdot,1,3}$ & 0.0004 & 0.0000 & 0.0016 & 0.0008 & 0.0001 & 0.0030 & 0.0004 & 0.0000 & 0.0017\\ 
  $p_{\cdot,1,4}$ & 0.0000 & 0.0000 & 0.0000 & 0.0000 & 0.0000 & 0.0001 & 0.0000 & 0.0000 & 0.0000\\ 
  $p_{\cdot,1,5}$ & 0.0000 & 0.0000 & 0.0000 & 0.0000 & 0.0000 & 0.0001 & 0.0000 & 0.0000 & 0.0000\\ 
   \hline
   \end{tabular}
   \caption{Sensitivity analysis of the posterior parameters with different priors. Scenario 1 sets $\bm{p}_{i1} \sim Dirichlet(2.5, 38.5, 0.5, 0.5, 0.5)$. Scenario 2 sets the growth parameter $\alpha_c \sim N(0,1), c=2,\dots, 5$. Scenario 3 combines both changes to the priors from scenario 1 and 2. }\label{tab:sensitivity}
 \end{table}

\subsection{Local structural impacts of common mutations}\label{sec:resultstruct}

For each of the nine segments identified in Figure \ref{fig:ref3D}, we ran the SMC and DiSGro algorithms to sample conformations for the reference sequence and the mutated sequence.  The specific segments, including the starting and ending positions, are shown in the first three columns of Table \ref{tab:rmsd}.  Note that each of the segments considered contains a single mutation, for example, the length 15 segment 923-937 sampled for A930V does not overlap with 929-943 for D936Y since they occur in different clusters.  For DiSGro \citep{tang2014fast}, we used the program from the authors to generate 100000 conformations and kept the 5000 with the lowest energies as the representatives.  For SMC \citep{wong2018exploring}, we ran the algorithm with 60000 particles as in \citet{wong2020assessing}, and also kept the 5000 with the lowest energies as the representatives.

\begin{table}[htbp]
\begin{center}
\begin{tabular}{ccccccc}
\hline
 & & &  \multicolumn{2}{c}{\text{RMSD}$_R$} &  \multicolumn{2}{c}{\text{RMSD}$_{RM}$} \\
Cluster & Mutation & Sampled segment &   SMC & DiSGro  &   SMC & DiSGro \\
\hline
I,III,IV,V & D614G & 607--621 & 2.98 & 2.56 & 1.99  & 1.93 \\
III,V & A222V & 215--229 &  1.19 & 1.66 & 1.04  & 1.63 \\
III & D574Y & 567--581 & 5.35  &  5.13 & 2.40 & 12.19\\
III & H655Y & 648--662 & 1.41 & 1.71 & 1.48   & 1.54 \\
IV & S477N & 470--484 & 3.76 & 14.71  & 3.14  & 13.05 \\
IV & T632N & 625--639 & 2.91 & 4.44 & 2.71  & 4.86\\
V & P272L & 265--279 & 0.64 & 0.95 & 0.92  & 1.90\\
IV & A930V & 923--937 & 0.89 & 0.76 & 1.15  & 1.04\\
I & D936Y & 929--943 & 4.17  & 1.72 & 6.33 & 2.44\\
\hline
\end{tabular}
\caption{Results for the lowest energy conformations sampled by the SMC and DiSGro algorithms on the reference and mutated sequence segments.  The RMSD$_R$ columns calculate the RMSD between the reference 3-D structure and the lowest energy conformation sampled for the reference sequence, thus measuring the accuracies of the algorithms for reconstructing each of these segments of the S-protein.  The RMSD$_{RM}$ columns calculate the RMSD between the lowest energy conformations sampled for the reference sequence and mutated sequence, thus measuring the extent to which the location of the energy mode may have shifted in 3-D space as a result of mutation.} \label{tab:rmsd}
\end{center}
\end{table}

To summarize the distributions of these sampled low-energy conformations in 3-D space, we used a similar metric as in \citet{wong2020assessing}, by computing all pairwise RMSDs between conformations.  For each individual segment, let $x^{(1)}_R, \ldots, x^{(5000)}_R$ denote the 5000 conformations sampled for the reference sequence, and $x^{(1)}_M, \ldots, x^{(5000)}_M$ denote the 5000 conformations sampled for the mutated sequence.  Then we computed three sets of RMSDs, defined via
\begin{eqnarray}
d_{RR} &\doteq&  \left\{ \text{RMSD}(x^{(k)}_R, x^{(l)}_R) \right\}  \text{ for  } k,l \in \{1,2, \ldots, 5000\} \text{ such that } k \ne l, \nonumber \\
d_{MM} &\doteq&  \left\{ \text{RMSD}(x^{(k)}_M, x^{(l)}_M) \right\}  \text{ for  } k,l \in \{1,2, \ldots, 5000\} \text{ such that } k \ne l, \nonumber \\
\text{and~~~} d_{RM} &\doteq&  \left\{ \text{RMSD}(x^{(k)}_R, x^{(l)}_M) \right\}  \text{ for all } k,l \in \{1,2, \ldots, 5000\}.  \nonumber
\end{eqnarray}
Thus the set $d_{RR}$ approximately represents the distribution obtained by repeatedly sampling two random low-energy conformations from the reference sequence and computing the RMSD between those conformations; an analogous interpretation applies to $d_{MM}$ for the mutated sequence.  Meanwhile, $d_{RM}$ considers pairwise RMSDs between one random conformation from the reference sequence and one random conformation from the mutated sequence.  Visual differences between the histograms of $d_{RR}$ and $d_{RM}$ (or $d_{MM}$ and $d_{RM}$) would thus suggest that the low-energy conformations for the reference and mutated sequences lie in distinct regions of 3-D space.

\begin{figure}[htbp]
\begin{center}
\includegraphics[width = 0.9\textwidth]{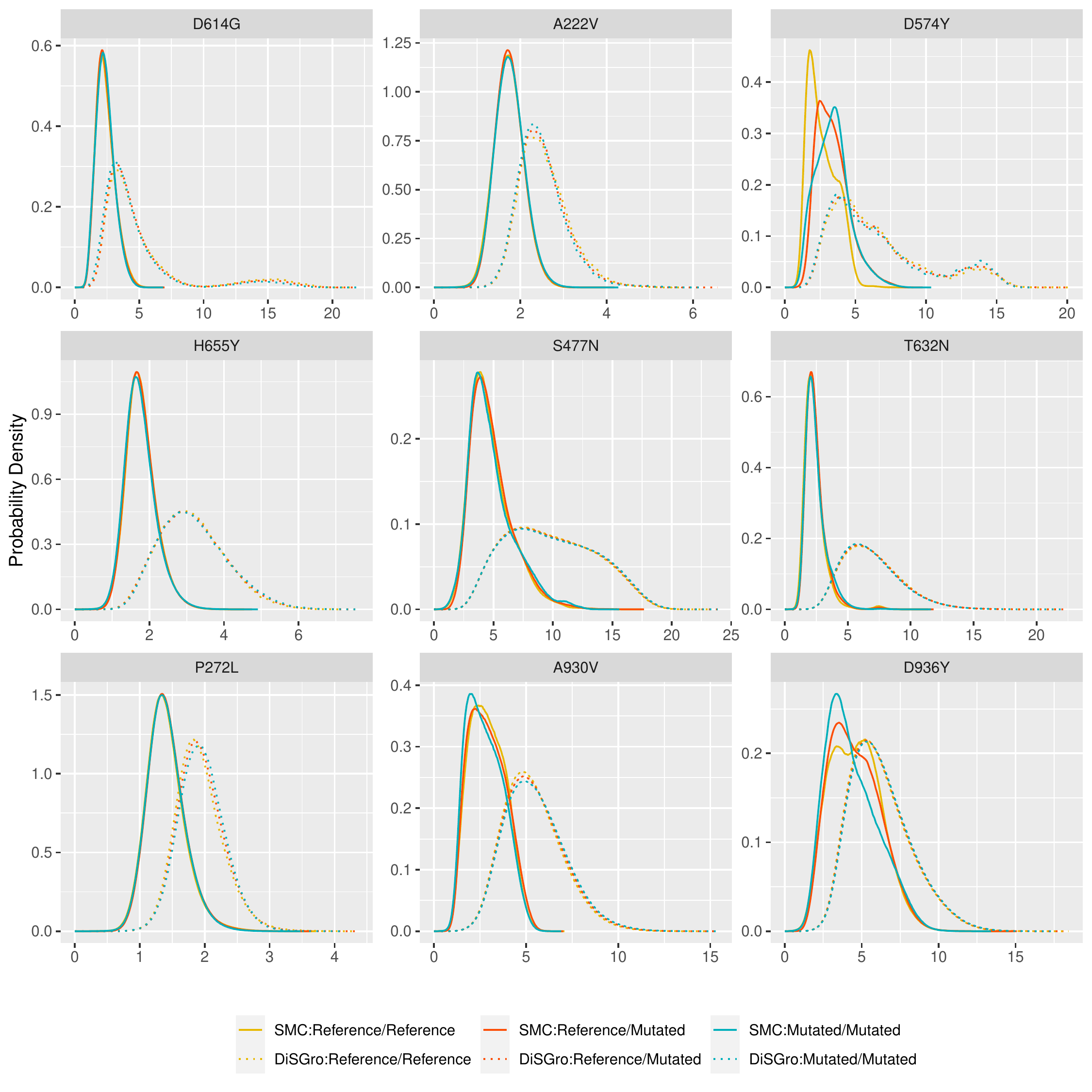}
\caption{Probability densities of the pairwise RMSD distributions  $d_{RR}$ (Reference/Reference), $d_{RM}$ (Reference/Mutated), and $d_{MM}$ (Mutated/Mutated) for each of the nine segments in Table \ref{tab:rmsd} based on the sampled conformations from SMC (solid lines) and DiSGro (dotted lines).  The $x$-axes are RMSDs in units of Angstroms.} \label{fig:pairrmsd}
\end{center}
\end{figure}

Plots for $d_{RR}$, $d_{RM}$, and $d_{MM}$ for each of the nine segments are shown in the panels of Figure \ref{fig:pairrmsd}, normalized to be probability densities and smoothed via kernel density estimation \citep{botev2010kernel}, labeled as Reference/Reference, Reference/Mutated, and Mutated/Mutated respectively.  These RMSD distributions are largely indistinguishable for most segments, suggesting that there is little discernible impact to the local conformational space of the protein as a result of the mutation, regardless of which sampling algorithm is used.  Three of the nine panels do have some visible differences between these RMSD distributions when sampled via the SMC algorithm:  D574Y, A930V, and D936Y.  In each case, the Reference/Mutated probability density visually appears as a compromise between the Reference/Reference and Mutated/Mutated densities, which is sensible.

In addition to the overall RMSD distributions, we may also specifically examine the lowest energy conformation, as is often done in  protein structure prediction applications. First, we considered conformations sampled for the reference sequences, where the true structure is known from the PDB.  For both DiSGro and SMC methods, we computed the RMSD between the true structure and the lowest energy conformation sampled by the algorithm.  These results are shown in the RMSD$_R$ columns of Table \ref{tab:rmsd}, which may be interpreted as the prediction accuracy if the algorithms are tasked with reconstructing the 3-D structure for each of these segments.  Overall, these results show reasonable accuracies, with the segments 567--581 and 470--484 being the most difficult to predict correctly for both algorithms.  Second, we considered conformations sampled for the mutated sequences, again taking the lowest energy conformation sampled by both algorithms.  Here, the true structures are unknown (except for D614G) so prediction accuracy cannot be assessed in general.  Thus, in the RMSD$_{RM}$ columns of Table \ref{tab:rmsd}, we instead compute the RMSD between the lowest energy sampled conformations for the reference and mutated sequences, as a way to quantify whether location of the mode of the energy distribution (as approximated by the samples) has shifted significantly after mutation.  Here, both algorithms agree in predicting that mutations D614G, A222V, H655Y, P272L, and A930V result in relatively small local 3-D structural changes (RMSD $< 2$) in the lowest energy conformation, while larger local structural changes are predicted by one or both algorithms for D574Y, S477N, T632N, and D936Y.

For D614G, we may validate the sampling results as this mutation has been studied in the laboratory with a determined 3-D structure in the PDB  \citep[accession code 6XS6,][]{yurkovetskiy2020structural}.  The actual RMSD between the reference structure (6XM0) and 6XS6 computed over the positions 607-620 corresponding to the sampled segment is 0.38 (coordinates for position 621 are missing in 6XS6); in contrast, the RMSD when computed over the larger structural unit from positions 531 to 620 is 2.61.  This result indicates that the local structural change as a result of the D614G mutation is indeed quite small, which is in agreement with the predictions of the sampling algorithms. The D614G mutation does however lead to more substantive global changes to the S-protein structure, which would be very difficult to predict computationally;  general protein structure prediction remains a highly challenging problem, despite recent progress \citep{kryshtafovych2019critical}.

\section{Discussion}

In this paper, we presented statistical approaches to tackle the challenges associated with the analysis of S-protein sequence and structure data. First, to better understand the evolution of S-protein sequences, we grouped the S-protein sequences into hierarchical clusters, and studied the spatial and temporal trend of these mutant clusters using a Bayesian hierarchical model. Second, we used sampling algorithms to investigate the possible changes in the local 3-D structure of the S-protein in the segments where the most frequent mutations occurred.

Based on our model estimates, we found that on average the reference sequence and its closely-related variants will diminish, while variants with the co-occurring mutations of D614G together with S477N or A222V tend to increase most strongly in prevalence over time. Our estimates of trend not only examined individual mutations as was analyzed in \cite{Korber}, but also captured the prevalence of some co-occurring mutations that have so far received limited attention in the literature.  Nonetheless, our findings on the reference sequence do align with \cite{Korber}, where the authors showed that a transition of position 614 from D to G occurred in many regions around the world with varying levels of statistical significance. Our estimates of S477N and A222V are in agreement with the trends observed by the Los Alamos National Laboratory, and A222V in particular is also consistent with \cite{Hodcroft} where the authors reported its presence in the majority of sequences in Europe by the fall of 2020. In addition, we found spatial dependence in COVID-19 transmission across country boundaries to be low in general, but higher within Europe. This could be related to their relatively loose travel policies within EU members during COVID-19 \citep{euro}.  Finally, a useful feature of our Bayesian approach is the ability to make projections of cluster proportions in countries where data is scarce or missing.

The result of the sequence analysis suggests potential fitness advantages or higher infectiousness for the co-occurring mutations D614G + S477N and D614G + A222V. In reality, while higher infectiousness may fully explain their growth in prevalence, other epidemiological factors may also play a role, for example, the characteristics of the infected population and the founder effect \citep{Korber}. Although \cite{LiQ} confirmed that D614G combined with other mutations (e.g., L5F, V341I, K458R, etc.) are more infectious than the reference sequence, the infectivities of D614G + S477N or D614G + A222V have not yet been mentioned and examined. Therefore, further experimental evidence is needed to confirm the increased infectivity of these co-occurrent mutations.

Having identified differences in the relative growth rates of the five mutant clusters considered, we examined whether the most common sequence mutations in these clusters were associated with changes in the 3-D structure of the protein near the mutation location.  Based on two different sampling algorithms, we conclude that evidence for large local structure changes is generally weak.  This computational result is consistent with the ground truth in the PDB for the one mutation (D614G) that has been studied in the laboratory thus far.  For the mutations S477N and T632N which are associated with cluster IV, both algorithms agree that a shift in the local 3-D conformation with lowest energy might be possible (with change in RMSD greater than $\sim$3).  Since protein structure determination experiments cannot keep pace with genome sequencing, we anticipate that computational approaches will continue to play an important role in understanding the possible structural impact of mutations.

Overall, S-protein sequence and structure datasets are a rich source of information that further research efforts can leverage for better understanding COVID-19, and we list some examples.  First, the sequence data could be expanded to include other data sources; the sequences used in this paper were collected from GISAID only where the sequence deposition rates from some countries is low.  Second, the sequence data could be combined with data on COVID-19 testing and case counts to estimate the actual prevalence of mutant clusters in different countries, in addition to their proportions.  Third, many laboratories have separately contributed S-protein structures to the PDB and these could be further analyzed to quantify uncertainties associated with structure determination efforts.


\section*{Acknowledgements}
This work was partially supported by a Discovery Grant from the Natural Sciences and Engineering Research Council of Canada.

\bibliographystyle{apalike} 
\bibliography{covid19}{}

\end{document}